\documentclass{iau}
\usepackage{graphicx}

\title[Recent and future instruments for DIB research] 
{The promise of recent and future observatories and instruments}

\author[Lex Kaper]   
{Lex Kaper$^1$}

\affiliation{$^1$Astronomical Institute Anton Pannekoek, University of Amsterdam, \\ Science Park 904,
1098 XH, Amsterdam, the Netherlands \\ email: {\tt L.Kaper@uva.nl} }

\pubyear{2013}
\volume{297}  
\pagerange{119--126}
\jname{The Diffuse Interstellar Bands}
\editors{Jan Cami \& Nick Cox, eds.}
\begin{document}

\maketitle

\begin{abstract}
  The identification of the carrier(s) of diffuse interstellar bands
  (DIBs) is one of the oldest mysteries in stellar spectroscopy. With
  the advent of 8-10m-class telescopes substantial progress has been
  made in measuring the properties of DIBs in the optical and
  near-infrared wavelength domain, not only in the Galaxy, but also in
  different environments encountered in Local Group galaxies and
  beyond. Still, the DIB carriers have remained unidentified. The
  coming decade will witness the development of extremely large
  telescopes (GMT, TMT and E-ELT) and their instrumentation. In this
  overview I will highlight the current instrumentation plan of these
  future observatories, emphasizing their potential role in solving
  the enigma of the DIBs.  
\end{abstract}

\firstsection 
\section{Introduction}

At a symposium in Noordwijkerhout, close to Middelburg where Hans
Lipperhey worked as an optician, and to The Hague where he
demonstrated the telescope to Prince Maurits, the Stadholder of
Zealand and Holland, and (unsuccessfully) requested patent on his
telescope, it seems appropriate to briefly memorize the telescope as a
Dutch invention. The invention of the telescope has always been a
matter of much controversy, but a recent publication of a newsletter
distributed in 1608 (\cite{Zoomers08}) provides supporting evidence
that the telescope, as a tool to study the stars, was invented by
Lipperhey.

The newsletter {\it Ambassades du Roy de Siam envoy\'{e} a
  l'Excellence du Prince Maurice, arriv\'{e} \`{a} La Haye le 10
  Septembre 1608} reports on three important events: the first visit
of a Siamese diplomatic mission to Europe, the peace negotiations with
Spain, {\it and} the demonstration of the newly invented telescope by
Hans Lipperhey at The Hague\footnote{The newsletter was intended to
  bring the news of the arrival of the Siamese embassy, leaving two
  and a half blank pages which were used to add the news of the
  telescope. Only three remaining copies of the newsletter are
  currently known.}, almost one year before Galileo Galilei was able
to discover the moons of Jupiter with an improved version of the
telescope. Since 1568, the Dutch Republic was in a state of war with
the Spanish empire of King Philips {\sc ii}. This ``Eighty Years War''
would last until 1648, but was interrupted by a Twelve Years Truce
from 1609--1621. In 1608, the commander-in-chief of the Spanish
forces, Ambrogio de Spinola, was in The Hague to represent Spain in
the peace negotiations. In those years, the Dutch East India Company
(VOC), the world's first multinational, was setting up a trade mission
in the small Malay state Patani (today a southern province of
Thailand), considered by the Dutch as the entrance to Siam and to
China. The Portuguese had started the rumour that the Dutch were
buccaneers without a country of their own, and so the Dutch welcomed
the initiative of the Siamese King Ekathotsarot to send a mission to
Holland.

The newsletter reports that ``A few days before the departure of
Spinola from The Hague, an optician from Middelburg offered a few
glasses to His Excellence ... with which it is possible to see the
windows of the church in Leiden. [...] the mentioned glasses are very
usefull to inspect objects at a distance of a mile and further, as if
they are very nearby. Even the stars that are usually invisible due to
their small size and our poor sight, can be seen with this
instrument.''.  Late November 1608 the newsletter had reached the
Italian philosopher Paolo Sarpi in Venice, and although his
correspondence seems to minimize its importance, in 1609 Sarpi became
the key intermediary in the contacts established by his friend Galileo
Galilei with the Venitian government regarding the telescope. The
newsletter is very specific about Hans Lipperhey being the inventor
and demonstrator of the telescope in 1608. One has to keep in mind,
however, that both Jacob Metius of Alkmaar and Sacharias Jansen of
Middelburg also claimed the invention of the instrument. This more
general knowledge of the instrument and its construction was the main
reason Lipperhey did not receive the patent he requested.

\begin{figure}[t]
\begin{center}
 \includegraphics[width=\textwidth]{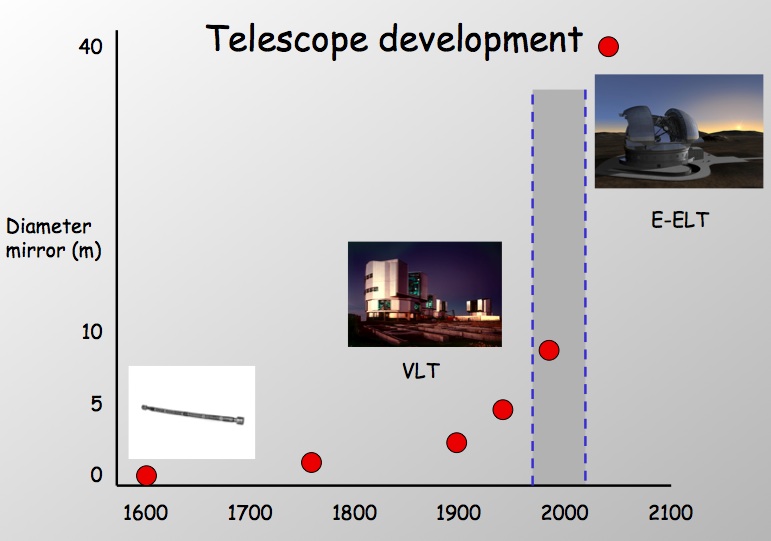} 
 \caption{Increase in telescope primary mirror (c.q. lens) diameter
   since its discovery in 1608.}
   \label{fig_mirror}
\end{center}
\end{figure}

Since 1608 the diameter of the primary mirror/lens has grown
exponentially with time (Fig.~\ref{fig_mirror}). Sir William Herschel,
discoverer of Uranus in 1780, built fantastic telescopes with
apertures exceeding a meter that were used on several sites. Early
$20^{\rm th}$ century the first telescope with a 2.5~m mirror was
constructed, the {\it Hooker} telescope of Mount Wilson Observatory
used by, among others, Edwin Hubble. With the construction of the 5m
{\it Hale} telescope in 1948, the maximum size of the conventional
reflector had been reached. Just before the millenium change,
alternative designs (segmented mirror, supported meniscus mirror)
resulted in the first 8--10m class telescopes (e.g. the {\it Keck}
telescopes and the ESO {\it Very Large Telescope}). Although we are
still exploring their scientific potential, the next generation of
30~m telescopes ({\it Giant Magellan Telescope} (GMT), the {\it Thirty
  Meter Telescope} (TMT) and the ESO {\it Extremely Large Telescope}
(E-ELT)) is planned to see first light in the next decade. 

Obviously, it is not only the telescope diameter that counts; also the
available suit of instruments determines for a large part the
scientific power of the observing facility. For the study of the
diffuse interstellar bands (DIBs), the identification of which termed
``the oldest mystery in stellar spectroscopy'', the development of
wide-band, high spectral resolution spectrographs (and
spectropolarimeters) is of special relevance. In the following
sections we present a (biased) view on current and future
instrumentation relevant for DIB research, with emphasis on the
current 8--10m telescopes and on the future ELTs.

\section{A biased view on current (and planned) instrumentation
  relevant for DIB research}

\subsection{DIBs in the Local Group and beyond}

The prominent DIBs at 5780 and 5797~\AA\ were first encountered in
B-star spectra and suspected to be of interstellar origin almost a
century ago (\cite{Heger22}, \cite{Herbig75}). Since then, several
hundred DIBs have been registered in various Galactic sightlines
(e.g., \cite{Tuairisg00}, \cite{Cox05}, \cite{Hobbs09},
\cite{VanLoon09}, \cite{Vos11}), while the nature of their carrier(s)
has, so far, remained illusive (\cite{Herbig95}, \cite{Sarre06}). 

Taking advantage of the increasing light collecting power of
telescopes, spectrographs and detectors, it became possible to search
for DIBs in extragalactic environments, and to study their behaviour
as a function of metallicity, extinction, interstellar radiation
field, etc. The first detections of DIBs in Local Group galaxies
focussed on the presence of the 4430~\AA\ DIB in the Magellanic Clouds
in the sixties and seventies (cf.\ review by \cite{Snow02}). The first
clear detections of LMC DIBs were obtained in the sightline towards
SN1987A (\cite{Vladilo87}). A more systematic study of DIBs in the
Magellanic Clouds was initiated by \cite{Ehrenfreund02} requiring
8+m-class telescopes for high-resolution spectroscopy towards reddened
OB stars: see \cite{Cox06}, \cite{Welty06}, \cite{Cox07},
\cite{VanLoon13}, the latter study exploring the capacities of a
multi-object spectrograph. Significantly more challenging regarding
DIB detection are the about ten times more distant spiral galaxies M31
(\cite{Cordiner11}) and M33 (\cite{Cordiner08}).

Extremely bright point sources (supernovae, gamma-ray burst
afterglows, quasars) provide the opportunity to search for DIBs beyond
the Local Group. DIBs are detected in high-resolution spectra of
supernovae hosted in several spiral galaxies: NGC~1448 ($d \sim
17$~Mpc, \cite{Sollerman05}), M~100 ($d \sim 16$~Mpc,
\cite{CoxPatat08}), and NGC~2770 ($d \sim 26$~Mpc, \cite{Thoene09}),
and in some starburst galaxies (\cite{Heckmann00}). Damped
Lyman$\alpha$ (DLA) systems in quasar spectra have also been inspected
for the presence of DIBs. Although these systems are gas rich ($N_{\rm
  H} > 2 \times 10^{20}$~cm$^{-2}$), only a few DLAs (12 out of 68,
\cite{Noterdaeme08}) show evidence for a molecular content (mainly
H$_{2}$, a few including HD and/or CO). \cite{Junkkarinen04} reported
the 4428~\AA\ DIB in a moderate-redshift DLA (towards the BL~Lac
object AO~0235+164 at a redshift $z = 0.52$); the spectrum also
includes a strong 2175~\AA\ feature. \cite{York06} also detected the
5705 and 5780~\AA\ DIBs in this sightline. A systematic search for
DIBs in another 6 DLAs resulted in upper limits only
(\cite{Lawton08}). \cite{Ellison08} reported another detection of the
5780~\AA\ DIB in the DLA and Ca~{\sc ii} absorber towards J0013--0024
at $z = 0.1556$.

Since the discovery of the first gamma-ray burst (GRB) optical
afterglow in 1997 (\cite{VanParadijs97}), about two hundred GRB
afterglows have been followed-up spectroscopically (e.g.,
\cite{Fynbo09}). The GRB host galaxy often produces a DLA from which
the redshift of the GRB can be determined (current record holder is
GRB090423 at $z=8.2$, \cite{Tanvir09}). The dust and metal content of
the GRB host galaxies is very low; up to now, no DIBs have been
detected in GRB hosts (the most prominent DIBs will be redshifted into
the near-infrared wavelength domain). The 2175~\AA\ extinction bump
has been detected in only four GRB host galaxies
(\cite{Zafar12}). Recently, \cite{Kruhler13} discovered H$_{2}$ (also
vibrationally excited H$_{2}$ due to the GRB) in the DLA towards
GRB120815A at $z=2.36$, but no DIBs were encountered (EW$_{\lambda
  6284} < 0.6$~\AA).

\subsection{Current (and upcoming) instrumentation}

The previous section (and other authors in these proceedings) report
on DIB observations carried out with medium- to high-resolution
spectrographs mounted at 2--10m-class telescopes. Excellent DIB
spectra can be obtained with the high-resolution spectrograph FEROS on
the ESO 2.2m telescope, UVES on the ESO {\it Very Large Telescope}
(VLT), HARPS on the ESO 3.6m telescope, MIKE on the {\it Magellan
  Telescope}, UCLES on the {\it Anglo-Australian Telescope}, ARCES on
the ARC 3.5m at Apache Point Observatory, HDS on the {\it Subaru
  Telescope}, HIRES on the {\it Keck Telescope}, etc. DIBs may also
produce linear and circular polarisation, although so far no
polarisation signal has been detected in DIB spectra
(\cite{Coxpol}). High-quality spectropolarimetry can be obtained on
relatively bright targets with, among others, ESPaDOnS on the CFHT,
NARVAL on the 2-m telescope {\it Bernard Lyot} and HARPSpol.

Several high-resolution spectrographs are currently being developed;
here we mention two running projects: PEPSI on the {\it Large
  Binocular Telescope} (LBT) and ESPRESSO for the combined focus of
the VLT. PEPSI is a fibre-fed high-resolution echelle
spectropolarimeter for the LBT (\cite{Strassmeier08}). It is designed
to use the two apertures of the LBT so that both circular and linear
polarisation signals can be simultaneously obtained. The
spectropolarimetric mode has a resolving power $R = $120,000; PEPSI
works in non-polarimetric mode at three different spectral
resolutions: $R = $32,000, 120,000 or 320,000. The wavelength range
from 390 -- 1050~nm can be covered in three exposures. The spectrograph is
contained inside a pressure- and temperature-stabilized
room. Commissioning of PEPSI on the telescope is scheduled in 2013.

ESPRESSO, the Echelle SPectropgraph for Rocky Exoplanet and Stable
Spectroscopic Observations, is a very stable spectrograph for the
combined Coud\'{e} focus of the VLT (\cite{Pepe10}). It can be
operated by either one of the VLT unit telescopes (UTs) or combine the light
of all four together. The wavelength coverage is limited, 380 --
686~nm, at a resolving power $R = 120,000$; with 4 UTs $R =
30,000$. The aimed instrumental radial-velocity precision is better
than 10~cm/s. The main scientific drivers for ESPRESSO are high-precision
radial-velocity measurements of solar-type stars to search for rocky
planets, the measurement of the variation of the physical constants,
and the analysis of the chemical composition of stars in nearby
galaxies. First light on the telescope is planned for 2016. 

\begin{figure}[t]
\begin{center}
 \includegraphics[width=12cm]{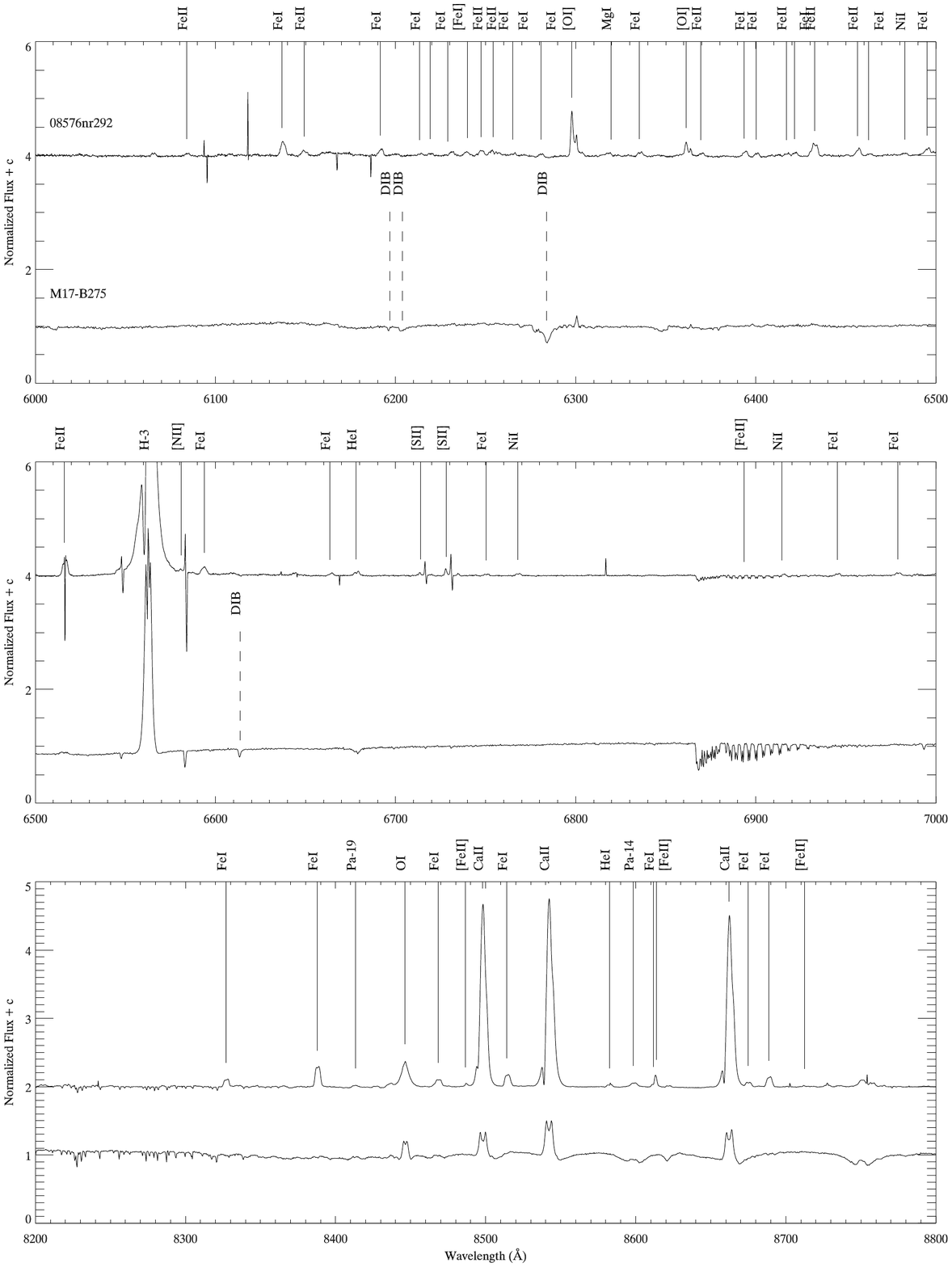} 
\caption{Part of the visual spectrum covered by VLT/X-shooter
  (6000--8000~\AA) of the reddened massive Young Stellar Objects 08576nr292 in
  RCW~36 ({\it top}, \cite{Ellerbroek11}, $A_{V} = 8$) and B275 in M17 ({\it
    bottom}, \cite{Ochsendorf11}, $A_{V} = 6$). The spectrum of 08576nr292 includes
  many emission lines produced by the circumstellar disk
  and jet. Also B275 exhibits (doubly-peaked) emission lines
  originating in a rotating disk; the spectrum, however, also includes
  photopsheric absorption lines allowing for accurate spectral
  classification: B6. Although both sightlines are severely reddened,
  the DIBs are relatively weak, especially towards 08576nr292 (figure
  adapted from \cite{Kaper11}).}
   \label{fig_mYSO}
\end{center}
\end{figure}

\begin{table}
 \begin{center}
   \caption{Obtaining spectra of O stars ($M_{V} \sim -6$) in the
     Local Group (and beyond) to detect DIBs will be possible up to
     the Virgo cluster with the ELTs. The current limit of
     VLT/X-shooter is $m_{V} \sim 22$ at a spectral resolving power $R
     \sim 8,000$, $S/N = 10$ in a 1~hour exposure
     (\cite{Vernet11}). The spatial resolution ($\theta = 1.22 \lambda
     / D$, with $D$ the diameter of the primary mirror) is important
     to avoid confusion, especially when moving out to more distant
     galaxies. The angular separation of two stars at a distance of
     0.3~pc from each other is listed in the last column.}
  \label{tab_Ostars}
  \begin{tabular}{lrcl} \hline
Target    & Distance   & $m_{V}$    & Separation 0.3~pc  \\
O star in  & (kpc)      &            & (arsec)            \\
\hline
SMC       & 68         & 13.2       & 1.0               \\
IC1613    & 730        & 18.3       & 0.09               \\
Cen A     & 4000       & 22.0       & 0.017             \\
Virgo cluster & 1600   & 25.0       & 0.004      \\
\hline
  \end{tabular}
 \end{center}
\end{table}

\subsection{Demonstrating the potential with VLT/X-shooter}

With the growth in telescope size we have witnessed the development
that traditionally Galactic studies have branched out into the
Magellanic Clouds and beyond. My expectation is that with the advent
of the ELTs extragalactic astronomy and astrophysics will merge. With
the ELTs it will be possible to take spectra of ``individual'' massive
stars in a representative sample of about 1500 galaxies out to 35~Mpc
(Tab.~\ref{tab_Ostars}).  To demonstrate this potential, we show a few
examples obtained with VLT/X-shooter, currently the most powerful
spectrograph in the world.

X-shooter is a medium-resolution ($R \sim 8000$), wide-band
echelle spectrograph covering the wavelength range 300--2500~nm in one
``shot'' (\cite{Vernet11}). Its three-arm design, with each arm
optimized for the covered wavelength band, and its location in the
Cassegrain focus of the VLT, make X-shooter a very efficient
instrument. It is one of the few spectrographs covering both the
optical and near-infrared wavelength range. This is interesting for
DIB research given the recent discovery of DIBs in the near infrared
(\cite{Geballe11}, Cox et al. in prep.). 

The formation process of massive stars is still poorly
understood. Their birth sites are deeply embedded in star-forming
regions and obscured by surrounding dust ($A_{V} \simeq 10 -
100$~mag). Optical to near-infrared spectra of massive Young
Stellar Objects (YSOs) provide information on the physical conditions
of the forming star(s), and on the nature of the accretion process
(disk/jet). We obtained X-shooter spectra of the YSO 08576nr292 in the
massive star-forming region RCW36 (\cite{Ellerbroek11}) revealing the
presence of a circumstellar disk and supersonic jets, demonstrating
that the object is still actively accreting (Fig.~\ref{fig_mYSO}). The
X-shooter spectrum of the mYSO B275 shows photospheric lines from
which the effective temperature and luminosity class are accurately
determined. It turns out that this B6 pre-main-sequence star has the
dimensions of a giant and is apparently still contracting towards the main
sequence (\cite{Ochsendorf11}). Fig.~\ref{fig_mYSO} also shows the
presence of DIBs in these reddened sightlines, though much weaker than
expected on the basis of their E(B--V) (\cite{Ellerbroek13}, and see
Oka, these proceedings).

\begin{figure}[t]
\begin{center}
 \hbox{
 \includegraphics[width=5.5cm]{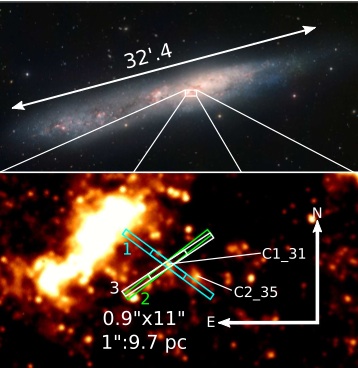} 
 \includegraphics[width=8cm]{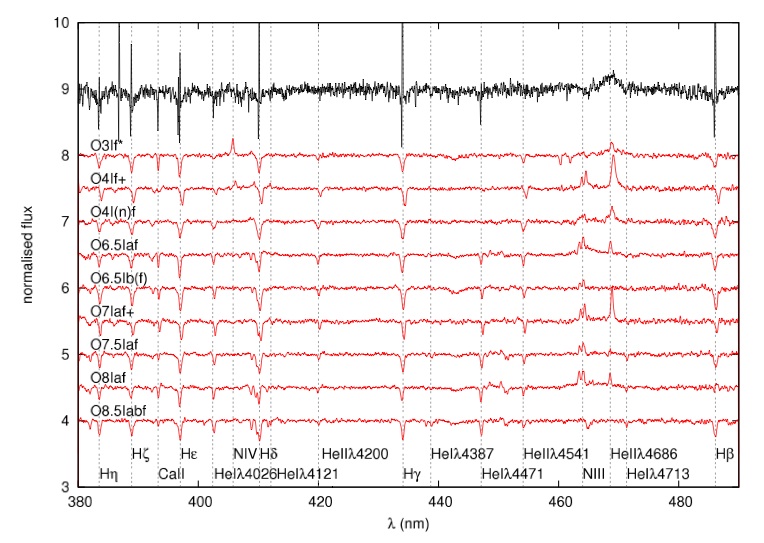}} 
\caption{{\it Left:} The top panel shows a B -- V -- H$\alpha$ image
  of NGC~55, a spiral galaxy in the Sculptor group ($d = 2$~Mpc). The
  bottom panel demonstrates the location of the slit of X-shooter
  centered on the massive star C1\_31. {\it Right:} The normalized
  sky-corrected X-shooter spectrum of NGC\_C1\_31 (top) compared to
  spectra of standard O supergiants ranging from spectral type O3 to
  O8.5. The C1\_31 sightline contains a hint of the 4430~\AA\ DIB,
  clearly present in the O6.5~Ib(f) comparison spectrum. Figures
  adopted from \cite{Hartoog12}.}
   \label{fig_NGC55}
\end{center}
\end{figure}

With 8--10m telescopes one can spectroscopically access the massive
star population in the Local Group and even beyond (e.g.,
\cite{Tramper11}, \cite{Bresolin13}). With X-shooter we obtained
spectra of a previously identified bright O-type star C1\_31 in
NGC~55, a spiral galaxy in the Sculptor group at a distance of $\sim
2$~Mpc (Fig.~\ref{fig_NGC55}). Analysis of both the stellar and
nebular spectrum yields that the source is a composite object, likely
a stellar cluster, which contains at least one hot WN star and a dozen
OB~supergiants (\cite{Hartoog12}). The C1\_31 sightline contains a
hint of the 4430~\AA\ DIB, clearly present in the O6.5~Ib(f)
comparison spectrum. This result shows both the potential of massive
star spectroscopy in more distant galaxies and the limitations due to
crowding and confusion.
 
\section{Extremely Large Telescopes}

The next decade is expected to revolutionize astronomy (again). With
the next generation Space Telescope, the NASA {\it James Webb Space
  Telescope} (JWST), the full array of the ESO {\it Atacama Large
  Millimeter Array} (ALMA), the {\it Square Kilometer Array} (SKA)
and, hopefully, a new large X-ray mission, the planned Extremely Large
Telescopes (ELT), with about a factor 10 more collecting area than the
current 8-10m-class telescopes, will complement the coverage of the
full electromagnetic spectrum. At economically difficult times, two
US-led projects ({\it Giant Magellan Telescope} and {\it Thirty Meter
  Telescope}) and one European-led project (ESO {\it Extremely Large
  Telescope}) will likely result in two ELTs in Chile and one in
Hawaii early next decade.

The scientific motivation to build an ELT is to address key
astronomical questions ranging from the formation of stars and planets
to unraveling the formation history of galaxies and large-scale
structure in the Universe.  In order to exploit not only the
collecting area of an ELT, but also its spatial resolution, a
sophisticated adaptive optics (AO) system is required employing
rapidly deformable mirrors. The AO system takes advantage of several
natural and/or artificial (laser guide star) point sources in the
field of view of the telescope to reconstruct the image plane that got
disturbed by the Earth's atmosphere.

\begin{table}
 \begin{center}
   \caption{Planned optical and/or near-infrared medium- to
     high-resolution spectrographs for the three ELTs: GMT, TMT and
     E-ELT.}
  \label{tab_ELTinst}
  \begin{tabular}{llccc} \hline
Instrument   & Function   & $\lambda$ range    & Resolving & Field  \\
             &            & ($\mu$m)           & power ($R =
             \frac{\lambda}{\triangle \lambda}$) & of view            \\ \hline
  \multicolumn{5}{c}{{\bf Giant Magellan Telescope (GMT)}} \\
\hline
G-CLEF       & High-res. spectrometer        & 0.35 -- 0.95
& 20,000 -- 100,000 & single object             \\
GMACS   & Optical MOS        & 0.36 -- 1.00       & 1500--4000, 10,000
& 40--50 arcmin$^{2}$          \\
NIRMOS     & Near-infrared MOS       & 0.9 -- 2.5     & 2700--5000   & 42 arcmin$^{2}$ 
\\ 
GMTNIRS    & AO-fed high-res. spectr. & 1.2 -- 5.0  & 50,000 --
100,000 & single object \\\hline
  \multicolumn{5}{c}{{\bf Thirty Meter Telescope (TMT)}} \\ \hline
WFOS & Wide-field opt. spectrometer   & 0.31 -- 1.1 & 1000 -- 8000
& 40.3 arcmin$^{2}$   \\
HROS & High-res. opt. spectrometer   & 0.31 -- 1.3  & 50,000 & 5 arcsec slit \\
NIRES & Near-IR AO-fed echelle & 1 -- 5 & 20,000 - 100,000 & 2 arcsec slit \\ \hline 
  \multicolumn{5}{c}{{\bf ESO Extremely Large Telescope (E-ELT)}} \\
\hline
HARMONI & Single-field IFU spectr. & 0.47 -- 2.45 & 4000 -- 20,000 & $1" \times 0.5"$ \\
CODEX & High-res. visual spectr. & 0.37 -- 0.71 & 135,000 & 0.82 arcsec \\
EAGLE & AO-assisted near-IR MOS & 0.8 -- 2.45 & 4000 -- 10,000 & $1.65" \times 1.65"$ \\
OPTIMOS-EVE & Opt-NIR fibre-fed MOS & 0.37 -- 1.7 & 5000 - 30,000 &
240 objects \\
            &                  &   &   & or IFUs \\
SIMPLE & NIR echelle spectr. & 0.8 -- 2.5 & 135,000 & 4 arcsec \\ \hline
  \end{tabular}
 \end{center}
\end{table}

\subsection{Giant Magellan Telescope (GMT)}

The GMT is the smallest of the ELTs, but has fewest reflections and
the largest field of view (20~arcmin). The primary mirror consists of
seven 8.4~m segments, resulting in a diameter of 25.4~m and an
effective area of a 21.9~m telescope. The secondary is also a
segmented mirror built up with seven 1.1~m segments. Las Campanas, in
the Atacama dessert in Chile, will be the site of the
GMT. Commissioning of the telescope is planned for 2019. The
instruments will all be located on a platform below the primary
mirror. The suite of candidate first generation GMT instruments
(\cite{Jacoby12}) contains two survey instruments, two high-resolution
spectrographs, and two AO imaging spectrometers
(Tab.~\ref{tab_ELTinst}): GMTNIRS, MIISE, HRCAM, NIRMOS and
GMACS. Likely a down-selection will be made after each of the
instrument concepts has been developed to the point of maturity. The
``Magellan philosophy'' is to keep the instruments simple, and to do a
few things well. MANIFEST, the planned robotic fiber feed facility
will provide a very efficient way to distribute the costy telescope
time to the different optical (fiber-fed) instruments.

\subsection{Thirty Meter Telescope (TMT)}

The TMT primary f1 mirror has a diameter of 30 meter consisting of 492
mirror segments of 1.45~m each; the field of view is 20~arcmin. It
will be a fully integrated adaptive optics telescope with a
multi-instrument Nasmyth mounted suite. Mauna Kea on Hawaii is the
preferred site for the TMT. The first scientific observations with the
TMT are planned in 2022. Eight instruments have been conceived to
attack the science problems (\cite{Simard10}); the three first light
instruments are the Wide Field Optical Spectrometer (WFOS), the
Infrared Imaging Spectrometer (IIS) and the Infrared Multi-object
Spectrometer. High-resolution spectrographs are planned to arrive at a
later stage.  

\subsection{ESO Extremely Large Telescope (E-ELT)}

\begin{figure}[t]
\begin{center}
 \includegraphics[width=\textwidth]{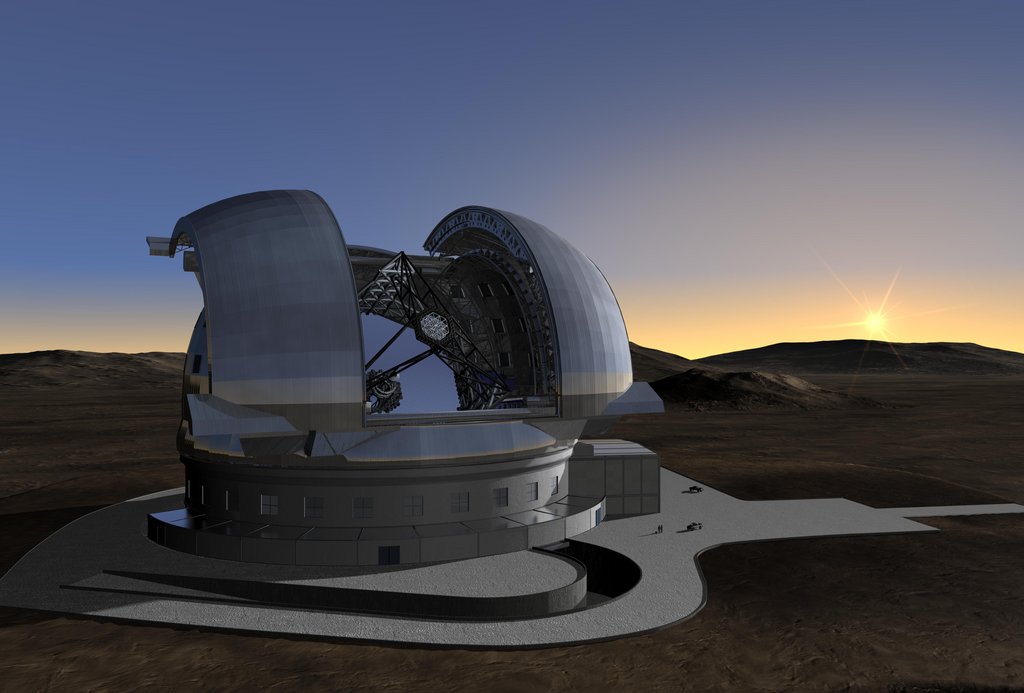} 
 \caption{Artist impression of the ESO {\it Extremely Large Telescope}
   to be built on Cerro Armazones in the Atacama dessert in Chile,
   about 20~km from Paranal. First light is scheduled for 2022.}
   \label{fig_E-ELT}
\end{center}
\end{figure}

Although the original plan started with a study of a 100~m diameter
telescope\footnote{It turns out that a telescope of such a diameter is
  required to image an Earth-like planet orbiting a nearby star, one
  of the key scientific motivations to build an ELT.}, the {\it
  Overwhelmingly Large Telescope} (OWL), the E-ELT design has shrunk
from a 42-m to a 39-m primary mirror to make sure that the project
fits within its budget envelope, and that the main scientific goals can
still be met. The E-ELT will be the largest optical/near-infrared
telescope in the world. The optical design of the E-ELT differs from
that of the other ELTs, as it includes adaptive mirrors in the
telescope. The novel five-mirror design should result in an
exceptional image quality with no significant aberrations in the
10-arcmin field of view (physically about two by two meter!). The
E-ELT is planned to be able to correct for the atmospheric
distortions from the start, providing images 16 times sharper than
those from the {\it Hubble Space Telescope}. The E-ELT's technical
first light is scheduled for 2022.

The first steps in developing the E-ELT instrumentation plan included
so-called phase-A studies of eight instrument concepts that were
completed in 2010 (\cite{Ramsay10}). Two instrument concepts have been
selected for first light: a diffraction-limited near-infrared imager
(MICADO) and a single-field near-infrared wide-band integral field
spectrograph (HARMONI). The third to fifth instrument will be a
mid-infrared imager and spectrometer (METIS), an optical/near-IR MOS,
and a high-resolution spectrograph. The EAGLE and OPTIMOS-EVE MOS
consortia are investigating the possibility to merge their respective
instrument concept into one, called MOSAIC (\cite{Evans13}), and aim
for the third slot on the E-ELT Nasmyth platform. GMT and TMT have
recognized the importance of a MOS for an ELT and have one or more
MOS concepts included in the first generation of instruments.

Astronomy has a bright future ahead; with a bit of luck we manage to
identify the DIB carrier(s) within 100~yr after the discovery of DIBs
and will be able to explore the information they
provide on the physical and chemical nature of the interstellar medium
in the Galaxy, the Local Group and beyond.

\end{document}